\documentstyle[twocolumn,prl,aps]{revtex}

\tighten

\begin{document}

\title{Expansion algorithm for the density matrix}

\author{Anders~M.~N.~Niklasson}

\address{
Theoretical Division, Los Alamos National Laboratory,
Los Alamos, NM 87545, USA}

\date{\today}
\maketitle

\begin{abstract}
{A purification algorithm for expanding the single-particle 
density matrix in terms of the Hamiltonian operator 
is proposed. The scheme works with a
predefined occupation and requires less than half the
number of matrix-matrix multiplications compared to
existing methods at low ($<10\%$) and high ($>90\%$) 
occupancy. The expansion can be used 
with a fixed chemical potential in which case
it is an asymmetric generalization of and a
substantial improvement over grand canonical McWeeny purification.
It is shown that the computational complexity, measured as
number of matrix multiplications, essentially is independent
of system size even for metallic materials with a vanishing band gap.
\\
}
\end{abstract}

\section{Introduction}
Theoretical predictions of material properties of complex systems
consisting of millions of atoms are often limited not by theory 
but by the calculational techniques. Recently there has been a large effort to develop 
ab-initio methods that computationally scale linearly with system size 
\cite{Goedecker_RMP_99}. The techniques may play an important role
in a broad spectrum of science such as molecular biology, materials science, chemistry and
nanotechnology.
Several of the linear scaling schemes are based on the single-particle density matrix 
that can be used in order to calculate the energies and densities that
occur in self-consistent field theories. The construction of the density matrix is
used as an alternative to solving an eigenvalue problem.
For large complex systems within a sparse matrix representation this approach 
can be performed more efficiently and instead of 
a cubic scaling the computational cost scales linearly with the system size 
\cite{Goedecker_RMP_99}. The matrix sparsity
is essential for the success of density matrix schemes.
For materials with a gap the real space 
representation of the density matrix is sparse \cite{Kohn59,Baer97,Stephan00}
due to a finite interaction length, which is usually referred to as nearsightedness \cite{Kohn96}. 
However, within other representations, such as a multi-resolution
wavelet basis or a group-renormalization representation, 
the density matrix is sparse also for metallic systems 
\cite{Beylkin91,White92,Goedecker99,Niklasson02}.

Most techniques for constructing the density 
matrix can be seen as a polynomial expansion of the density matrix $\rho_0$ 
in terms of the Hamiltonian operator $H$.
In an iterative approach this expansion can be formulated as
\begin{equation} \label{Purif}
\begin{array}{ll}
X_0 &= P_0( H) \\
X_{n} &= P_n( H, X_{n-1}), ~~~ n=1,2, \ldots\\
\rho_0 &= \lim_{n \rightarrow \infty} X_n.\\
\end{array}
\end{equation}
The projection polynomials $P_n(H,X_n)$ are chosen
to achieve a rapid convergence under the conditions of commutation, 
$[H,\rho_0] =0$, idempotency, $\rho_0^2 = \rho_0$,
and particle conservation, $Tr[\rho_0] = N_e$. They may either be chosen from
a constrained conjugate gradient minimization of the energy functional $Tr[\rho H]$
\cite{Kohn96,Sameh82,Li93,Carlsson95,Hernandez96,Daniels97,Yang97,Stephan98,Challacombe99,Haynes99,Bowler99,Daniels99},
or as a fast expansion of the step function $\theta(\mu I - H)=\rho_0$
centered at the chemical
potential $\mu$, or, for finite temperatures, the Fermi-Dirac distribution
\cite{Niklasson02,Bowler99,Daniels99,McWeeny60,Goedecker94,Kenney95,Palser98,Holas01}.
Each computational step consists of matrix-matrix additions, subtractions and multiplications. 
The problem is to find a rapidly convergent expansion that minimizes the number of
matrix-matrix multiplications, since these operations 
are the most time consuming \cite{Challacombe00M,Bowler01}. 

The efficiency of the different density-matrix schemes
varies depending on particular characteristics of the problem such as 
the existence of a band gap, a predefined chemical potential, 
filling factor, self-consistency cycles, thresholding, basis set and system size. 
In this letter we propose a purification algorithm for the construction of the density matrix
that is simple, general and rapidly convergent also for very large metallic systems.
The method works with a predefined occupation and does not need the input or
adjustment of the chemical potential. Only one previous purification strategy,
recently developed by Palser and Manolopoulos (PM) \cite{Palser98}, exists
for this important problem. By using a starting guess $X_0$ with 
the trace equal to the occupation number and thereafter performing trace-conserving 
spectral projections, $X_n$ converges to the correct density matrix without 
prior knowledge of the chemical potential. The PM scheme has an excellent
performance compared to other methods \cite{Palser98,Daniels99}.
However, due to the constraint of trace conservation the method is inefficient
at low and high partial occupancy. This is of great concern,
for example, when using a multi-resolution wavelet representation for metallic problems,
since the fractional filling in this case is low. The same
problem occurs with a minimal basis set at both high and low
occupancies. A simple general algorithm that avoids this particular problem, 
and still converges as or more rapidly, especially for very
large problems, would therefore be of great interest.

\section{Trace correcting purification}

The method we propose is based on the continuously increasing purification polynomials 
with stationary end points in $[0,1]$ 
\begin{equation} \label{Poly}
\left\{
\begin{array}{l}
P_{m}^{(a)}(x) = 1 - (1-x)^m\left[ 1+mx\right] \\
P_{m}^{(b)}(x) = x^m\left[ 1+m(1-x) \right].
\end{array} \right.
\end{equation}
Two examples of $P_{m}^{(a)}(x)$ and $P_{m}^{(b)}(x)$, for $m=1$ and $3$, 
are displayed in Fig.\ \ref{PolyFig}. It can be shown that any combination
of these polynomials in an iterative expansion 
converges to a step function for $x \in [0,1]$, i.e.\
\begin{equation}
\theta(x-\xi )= \ldots (P_{m}^{(a/b)}( P_{m}^{(a/b)}(x) )\ldots ,
\end{equation}
with the step $\xi \in [\beta_m,1-\beta_m]$. Here $\beta_m$ is the inflection point of 
$P_{m}^{(a)}(x)$, i.e.\ where $P_{m}^{(a)}(\beta_m) = \beta_m$, $0 < \beta_m < 1$,
and $(1-\beta_m)$ is the inflection point of $P_{m}^{(b)}(x)$. The convergence
towards a step function can be understood from the fact that for each new
iteration the new function will still be continuously increasing, but with an
in creasing number of vanishing derivatives at the end points. The asymmetry
in the number of vanishing derivatives determines the position of the step.
The choice $m=2$ corresponds to the McWeeny polynomial \cite{McWeeny60}.
In this symmetric case $\beta_2 = 1/2$, and a step can only be formed at $ \xi = 0.5$.
The occupation of an operator $X$ can be modified such that
\begin{equation}
\left\{
\begin{array}{ll}
Tr [ P_{m}^{(a)}(X) ] \geq Tr [ X ]; & \varepsilon (X) \in [\beta_m,1]\\
Tr [ P_{m}^{(b)}(X) ] \leq Tr [ X ]; & \varepsilon (X) \in [0,1-\beta_m],
\end{array}
\right.
\end{equation}
where $\varepsilon (X)$ are the eigenvalues of $X$.
For $m=1$ the reverse situation holds, with switched inequalities.
With $m \neq 2$ we can apply the polynomials of Eq.\ (\ref{Poly}) in the expansion
of the density matrix, Eq.\ (\ref{Purif}), such that
each step adjusts for the occupation of $X_n $. In this way 
an expansion is created that converges to the density matrix with the correct
occupation, i.e.\ $\rho_0 = \theta (\mu I - H)$ with $Tr[\rho_0] = N_e$, 
but without {\it a priori} knowledge of $\mu$. The algorithm (for $m > 2$) 
is given by this pseudocode:
\begin{equation} \label{Alg}
\begin{array}{l}
{\it function~ } \rho_0( H, N_e, {\it ErrorLimit})\\
{\it estimate~} \varepsilon_0(H),~\varepsilon_N(H) \\
X_0 = (1-2\beta_m)(\varepsilon_N I-H)/
(\varepsilon_N-\varepsilon_0) + \beta_m I\\
{\it while~ Error > ErrorLimit} \\
~~ {\it if}~ Tr [X_n ] - N_e < 0 \\
~~ ~~ X_{n+1} =  P_{m}^{(a)}(X_n)  \\
~~ {\it else}\\
~~ ~~ X_{n+1} = P_{m}^{(b)}(X_n) \\
~~ {\it end}\\
~~ {\it estimate~ Error} \\
{\it end}\\
\rho_0 = X_n ~ . \\
\end{array}
\end{equation}
For $m = 1$ the trace condition has to be reversed to $">"$.

The scheme can be described as follows:
First the Hamiltonian is normalized to an initial matrix $X_0$ with all its eigenvalues
$\varepsilon (X_0) \in [\beta_m,1-\beta_m]$.
The constants $\varepsilon_0$ and
$\varepsilon_N$ are the lowest and highest
eigenvalues of $H$, respectively. These can be approximated by, for example,
Gersgorin estimates or the Lanczos method, with only a small extra computational
cost \cite{Daniels99,Palser98}. A necessary criterion for convergence is that the
unknown chemical potential of the normalized initial matrix $\mu(X_0) \in [\beta_m,1-\beta_m]$. 
For intermediate occupancy, provided $\mu(X_0) \in [\beta_m,1-\beta_m]$,
$\beta_m$ can be set to zero in the starting guess.
This usually reduces the number of iterations by one or two steps.
The improvement has not been used in the present study.
After initializing $X_0$ the projections $P_{m}^{(a)}(X_n)$ or $P_{m}^{(b)}(X_n)$ are
performed, adjusting the occupancy and expanding a step function
at the same time. The iteration stops when some appropriate 
error estimate is less than a predefined error limit.
Note, that a high order expansion may lead to
too fast convergence, making the adjustment of 
the occupation impossible. We may also use combinations
with different values of $m$ as well as other asymmetric
purification polynomials. Any set of asymmetric continuously increasing polynomials
in $[0,1]$ with stationary points at $0$ and $1$ can be used equivalently.
The presented algorithm cannot handle problems with degenerate eigenstates at $\mu$. 
The algorithm would still converge, but to the
wrong density matrix, since the degenerate states would split due to numerical 
noise. This differs the presented trace correcting purification scheme 
from the PM scheme, which correctly can treat the case of degeneracy.

\section{Grand canonical purification}

Since any combination of the expansion polynomials in Eq.\ (\ref{Poly})
converges to a step function we can use a predefined fixed expansion 
combination of $\theta (\mu I - H)$. In this case $\mu$ must be known,
but the efficiency might be slightly improved compared to the
schemes working with a predefined occupancy.
For example, we may use the repetitions of the
combination $P_{\rm ex.}^{\rm GC}(x) = P_2^{(b)}(P_1^{(b)}(P_3^{(a)}(x)))$
or (for $m>1$) only $P_m^{(a)}(x)$, or only $P_m^{(b)}(x)$, in the
combination
\begin{equation}\label{GCPol}
P_m^{\rm GC}(x,\bar\mu) = \left\{
\begin{array}{ll}
P_{m}^{(a)}(x) & \bar\mu \geq 1/2 \\
P_{m}^{(b)}(x) & \bar\mu \leq 1/2,
\end{array} \right.
\end{equation}
where $\bar\mu$ is the normalized chemical potential,
$\bar \mu =(\mu -\varepsilon_0)(\varepsilon_N-\varepsilon_0)^{-1}$.
We can now, as in Eq.\ (\ref{Purif}),
perform the expansion using the fixed repeated combination of
$P_{m}^{(a)}(x)$ and $P_{m}^{(b)}(x)$ with the starting guess
\begin{equation} \label{StartG}
\begin{array}{ll}
X_0 = \alpha(\mu { I} - { H}) + \beta{ I},\\
\alpha = {\rm min} \left\{ \beta[\varepsilon_N - \mu]^{-1},
(1-\beta)[\mu - \varepsilon_0]^{-1} \right\}.
\end{array}
\end{equation}
The constant $\beta$ is here determined by the inflection
point of the repeated fixed polynomial combination, e.g.\  $\beta_m$
or $(1-\beta_m)$ for $P_m^{\rm GC}(x,\bar\mu)$. The approach 
can be seen as an asymmetric  generalization of the grand canonical
McWeeny purification scheme \cite{McWeeny60,Palser98,Holas01}. With 
$P_2^{\rm GC}(x,\bar\mu)$ they are equivalent. 
The method is directly related to matrix sign function
expansions \cite{Kenney95}. The matrix sign function expansion is equivalent
to the purification scheme via a trivial linear transform where the step function
expansion is performed between $-1$ and $+1$ in the interval $[-1,1]$.

If the chemical potential is unknown the density matrix 
may have to be recalculated with different values of $\mu$ until the occupation is correct.
However, since the density matrix can be described as a superposition of
outer products of the occupied eigenstates we can adjust the
occupation by adding or subtracting Hamiltonian eigenstates close to the
chemical potential. A few of these states can be calculated
efficiently, for example, using inverse power iterations \cite{Ipsen97}.
In this way the occupation can be adjusted without a complete 
recalculation of the density matrix with a new shifted chemical potential.
Moreover, in the case of a material with 
a gap we do not need to have a very precise prior knowledge of $\mu$
as long as the estimate is somewhere in the gap. The grand canonical approach, with a predefined 
fixed $\mu$, may thus be an efficient alternative 
to trace correcting purification in some special cases.

\section{Examples}

To illustrate the efficiency of the expansion techniques
we have constructed an $N \times N$ Hamiltonian test matrix $\bar H(i,j)$ with randomized
off-diagonal elements decaying as $|i-j|^{-2}$, and with a uniform distribution of 
eigenvalues in $[0,1]$.
Only the eigenvalue distribution of the Hamiltonian is of importance for the convergence.
With $\bar H$ we have that the occupation factor $\lambda = N_e/N = \bar \mu = \mu$ and
it is easy to compare grand canonical schemes with the trace correcting or trace conserving methods.

Figure \ref{Conv} shows the number of matrix
multiplications necessary to achieve an error $||X_n-\rho_0||_2 \leq 10^{-9}$
as a function of the filling factor $\lambda$ or chemical potential $\mu$.
The PM trace conserving purification scheme 
is slow at low and high occupancy (since the slope at the inflection point
tends to 1 as the inflection point approaches 0 or 1), 
whereas the new trace correcting expansion algorithm, with 
$m=1$ (P$_1$), $m=3$ (P$_3$), and $m=5$ (P$_5$), 
has an overall fast convergence. For example, at $10\%$ occupancy 
the new scheme with $m=3$ is about twice as fast compared to the PM scheme. 
There are essentially three reasons for the improved convergence: 
{\it i)} a more optimized ratio between the number of matrix multiplications
and the polynomial order, {\it ii)} a faster increase of the number of vanishing
derivatives at the stationary end points, and {\it iii)}, as will be shown below, 
a steeper slope at the inflection points, partly due to the asymmetry
of the purification polynomials. The generalized grand 
canonical expansion with $m=3$ (P$_3^{\rm GC}$) converges several steps
faster compared to the grand canonical McWeeny (McW) method for
the same reasons.

\section{Scaling}

By varying $N$, i.e. the size of the Hamiltonian test matrix, we may see how 
the number of matrix multiplications necessary for convergence
scales with system size, or equivalently with the inverse gap at the chemical 
potential $\Delta \varepsilon_{\mu} = 1/N$. The behavior is crucial
for very large systems, especially if we wish to construct an
expansion scheme that computationally scales linearly
with the system size for metallic materials with a vanishing band gap. 
Figure \ref{LinSc} displays the number of necessary matrix multiplications
as a function of $\ln (N)$. In the upper graph the results of McWeeny purification (McW)
and the trace conserving canonical purification (PM) are on top of each other. 
For this particular symmetrical case the two schemes are identical \cite{Palser98}.
The graph indicates a stepwise linear relationship 
between the number of matrix multiplications $M$ necessary for convergence
and the logarithm of the system size. The relation can be approximated by the linear formula 
\begin{equation}
\label{OrdN}
M(\mu , N)  = \alpha ( \mu ) + \kappa (\mu ) \ln (N).
\end{equation}
The least square fits of $M(\mu , N)$ are shown together with the values of $\kappa$.
The expansions P$_1$, P$_5$ and P$_3^{\rm GC}$ perform equally well or slightly
worse compared to P$_3$, and are not shown.
The slopes determine the efficiency for very large systems
and we find the best scaling for P$_3$.

The convergence is determined by the slowest converging eigenvalue $\gamma(X_0)$,
closest above or below the chemical potential of the normalized initial matrix $\mu(X_0)$. 
This particular eigenvalue should either converge to $1$ or $0$. Since
the purification polynomials are continuously increasing, preserving the
order of the eigenvalues, all other eigenvalues converge faster. 
In the case of grand canonical purification, with a uniform 
distribution of $N$ eigenvalues, $\gamma(N,X_0) \approx \beta \pm 1/(2N)$. 
By means of a linearization of the purification polynomial around $\beta$
it can be shown that $\gamma(N,X_0) \approx \gamma(kN,X_1)$, where $k$ is 
the derivative of the purification polynomial at the inflection point, e.g.\  
$k = P_m^{\rm GC}{'}(\beta_m,\bar\mu)$.
Thus, increasing the number of states by $\Delta N = N(k-1)$ and performing one
extra iteration leads to the same error in $\gamma$.
If ${\widetilde M}_\gamma$ is the number of multiplications necessary to achieve
a fixed error of $\gamma$ and $p$ is the number of matrix multiplication
in one iteration, we have that
\begin{equation}
 [{\widetilde M}_\gamma (N+\Delta N)-{\widetilde M}_\gamma (N) ]/(\Delta N) \approx p[N(k-1)]^{-1}.
\end{equation}
Let $M_\gamma$ be a continuous version of ${\widetilde M}_\gamma$ such that
\begin{equation}
dM_\gamma / dN = p \left[ N(k-1)\right]^{-1}.
\end{equation}
Integration gives
\begin{equation} \label{LinScA}
M_\gamma(N) = \alpha + \frac{p}{k-1} \ln (N),
\end{equation}
for some constant $\alpha$.
This approximate formula explains
the linear relation in Eq.\ (\ref{OrdN}) and is a useful measure
in optimizations of grand canonical expansions.
Purification polynomials should be optimized on the criterion of
matrix multiplications v.s. vanishing derivatives at the stationary fixed
points, and the slope $\kappa_{\rm est} = p(k-1)^{-1}$ 
as estimated in Eq.\ (\ref{LinScA}). 
For example, the purification ${P_3}^{\rm GC}(X)$ requires two matrix 
multiplications in each iteration, it has
three vanishing fixed point derivatives, and $\kappa_{\rm est} = 3.1$.
This should be compared to McWeeny purification that also requires
two matrix multiplications in each iteration, but with only two vanishing 
fixed point derivatives, and with a $\kappa_{\rm est} = 4.0$.

For materials with a band gap, the value $1/N$
should be replaced by the gap at the chemical potential $\Delta \varepsilon_{\mu}$.
In this case the scaling with the logarithm of the system size vanishes
and the number of necessary matrix multiplication for a predefined
convergence accuracy is constant. 

Notice that one may use different criteria for convergence such as the error per
state, per atom, or the total error. However, this has only a minor
effect on the number of necessary matrix multiplications
because of the very rapid rate of convergence close to idempotency, 
which, for example, is quadratic in the WcWeeny case.

\section{First principles performance}

To further illustrate the performance of the expansion scheme we 
show the result of an implementation in the MondoSCF suite of linear scaling 
self-consistent field programs \cite{Challacombe99,Challacombe97,Schwengler97,Challacombe00}.
Figure \ref{SCF} displays the number of necessary matrix multiplications 
for clusters and strings of Li atoms and for different molecules. 
In the case of Li clusters and strings of Li atoms the systems are metallic 
in the sense that the gap vanishes in the limit of infinite number of atoms. 
This is thus a good test to check the linear relationship between the
number of matrix multiplications and the logarithm of the inverse band gap,
i.e. the logarithm of the system size for metallic materials.
The P$_1$ scheme is efficient compared to the PM scheme, 
especially in the case of SiF$_4$ where the occupancy $\lambda$ is high.
This particular example illustrates the inefficiency of the PM scheme
at high and low occupancy which is avoided with the trace correcting algorithm.
The weak logarithmic dependence between computational cost and
system size for metallic systems, illustrated by the dashed lines in the figure, 
is also confirmed. Notice that an actual linear scaling is reached only if
the number of non zero elements of the density matrix grows
linearly with system size. This can generally not be achieved 
within a real-space representation for metallic systems. Instead, as
mentioned above, a multi-resolution wavelet basis or a group-renormalization 
approach has to be applied. However, the number of matrix multiplication
necessary for convergence should not be affected by a change of
representation since the vanishing gap around the chemical potential 
will remain the same regardless of the representation, given, in the case 
of a wavelets, via a biorthogonal transformation of the Hamiltonian 
\cite{Goedecker99,Niklasson02}.

The P$_1$ scheme implemented in the MondoSCF programs has two major
advantages compared to other schemes: {\it i)} It requires less memory
compared to higher order schemes since only second order polynomials are used and
intermediate matrix products does not have to be stored.
{\it ii)} It is less complex and only matrix squares has to be
calculated. A specially designed algorithm for matrix squares
can possibly be made more efficient than a general matrix
product algorithm.

\section{Discussion}

The expansion scheme and the convergence analysis illustrated and
argued for here provide a basis for the understanding of purification 
algorithms and their efficiency, and it shows that the
computational complexity, as measured in number of matrix multiplications,
essentially is independent of system size even for metallic
systems. This is in contrast to, for example, some conjugate gradient schemes, 
where in the worst case, the maximal number of iterations was shown to scale
as $N^c$, with $c$ varying between $1/3$ for insulators and $1$
for metals \cite{Annett95}. However, if the additional problem
of thresholding is included, which can be performed either via a finite
cut-off radius truncation, or via a numerical threshold, the computational
complexity as a function of system size, within some required numerical
accuracy, becomes far more difficult to analyze and practical experience
may be the only way to understand the efficiency.

In the alternative construction of the density matrix using a constrained
functional minimization, as devised by Li {\it et al.} \cite{Li93}, the
McWeeny purification is used to impose idempotency. The asymmetric polynomial
expansions proposed here may serve as a possible alternative.

\section{Summary}

In summary we have proposed an algorithm for expanding the single-particle 
density matrix in terms of the Hamiltonian that is simple,
general, and with a computational complexity essentially independent
of system size even for very large metallic systems with a vanishing band gap.
If the expansion is used together with a fixed chemical potential it was
shown to be an asymmetric generalization of the grand canonical McWeeny
purification. The algorithm is a substantial improvement 
of previous schemes and provides together with the presented 
convergence analysis a framework for the 
understanding and optimization of purification.

\section{Acknowledgement}

Discussions and support from Matt Challacombe and C.\ J. Tymczak are gratefully acknowledged.
I am also very thankful for editorial help from \ Eric Chisolm.

\begin{figure}
\caption \small{
Different projection polynomials for the adaptive expansion of the
step function in Eq.\ (\ref{Poly}).
\label{PolyFig}}
\end{figure}

\begin{figure}
\caption \small{
The number of matrix-matrix multiplications $M$ necessary
to achieve a convergence $||X_n - \rho_0 ||_2 \leq 10^{-9}$, as a function of 
the filling factor $\lambda$, or equivalently, the chemical potential $\mu$, 
for $\bar H$ with $N=100$.
PM corresponds to the result using the canonical trace conserving
purification scheme by Palser and Manolopoulos \cite{Palser98}. 
The open symbols P$_m$ show the result of the 
expansion algorithm, Eq.\ (\ref{Alg}).
The small squares indicate the result of the grand canonical WcWeeny purification (McW)
and $P_3^{\rm GC}(x,\bar\mu)$ in Eq.\ (\ref{GCPol}).
\label{Conv}}
\end{figure}

\begin{figure}
\caption \small{
The number of matrix-matrix multiplications $M$ necessary
to achieve a convergence $|E_n-E_0|/N \leq 10^{-9}$, as a function 
of $\ln (N)$ or equivalently $\ln (1/\Delta \varepsilon_\mu)$. 
Here $E_n = Tr [ \bar{H} X_n ]$, $N$ is the total number of states and
$\Delta \varepsilon_\mu$ is the gap at the chemical potential.
\label{LinSc}}
\end{figure}

\begin{figure}
\caption \small{
The number of matrix multiplications $M$ (after three
self-consistency cycles with STO-3G or STO-6G basis sets) necessary to achieve 
a convergence $|E_n-E_{n-1}| \leq 10^{-7}$ a.u., as a function of 
the logarithm of the number of atoms. Occupation $\lambda \in [0.3,0.7]$
except for NiF$_4$. For the Li strings and clusters, as indicated by the dashed lines, 
the gap is vanishing, i.e. metallic, in the limit ``Number of atoms'' $\rightarrow \infty$.
\label{SCF}}
\end{figure}

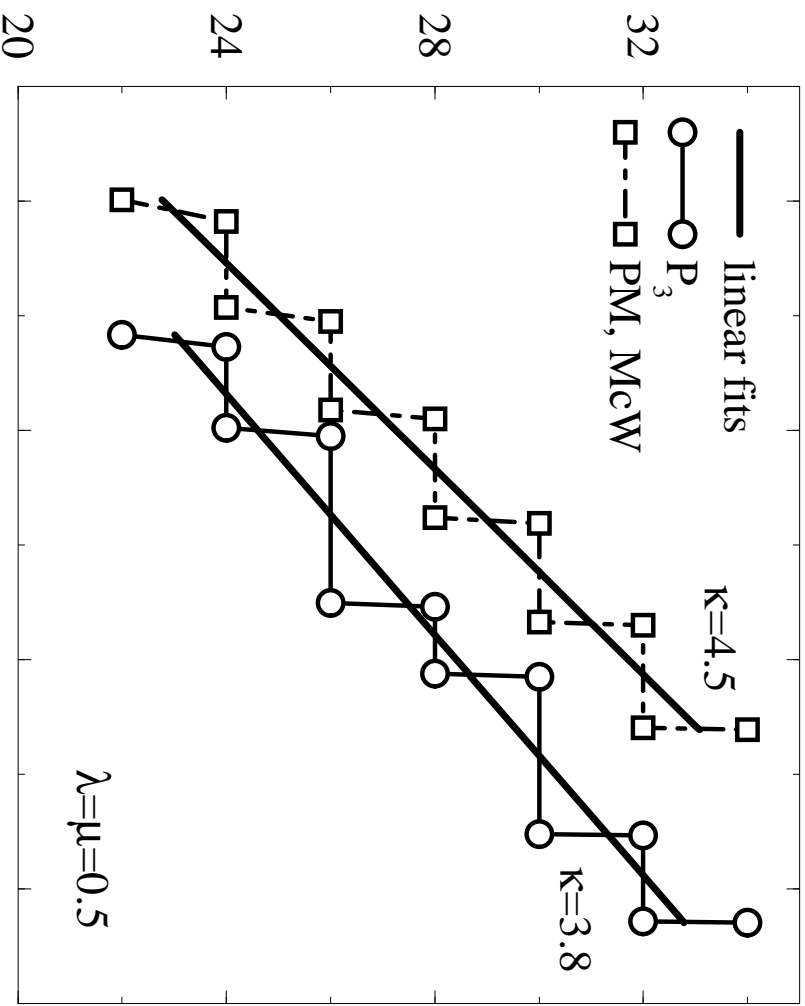
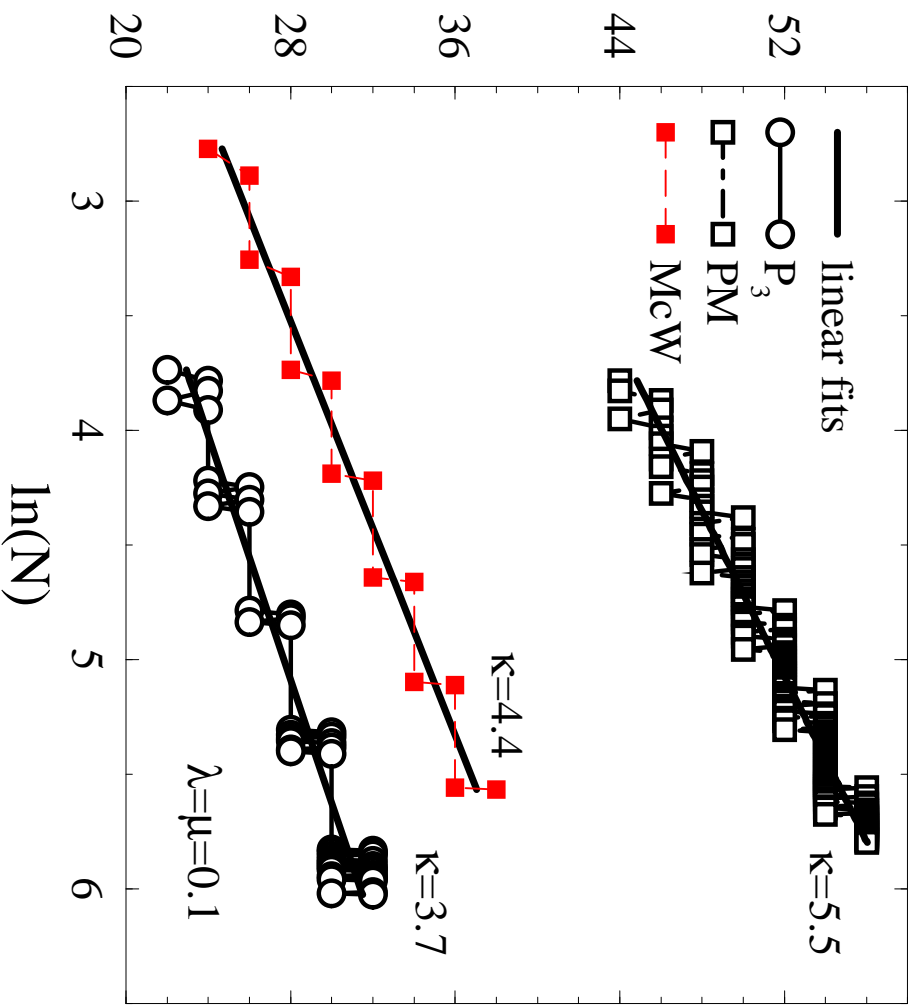
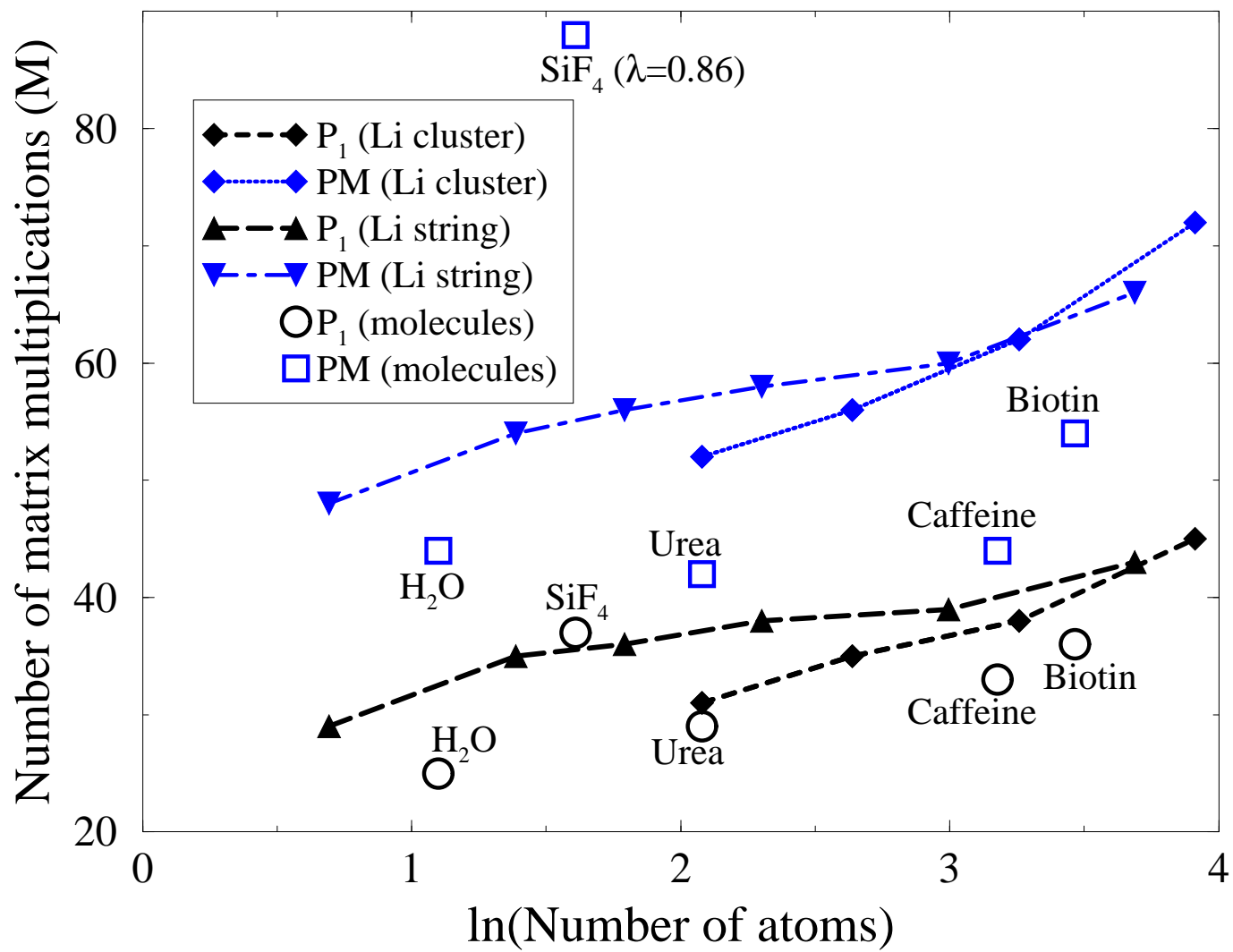
\end{document}